\documentclass[aps,prl,twocolumn,superscriptaddress,nofootinbib,showpacs]{revtex4}
\usepackage{slashed}
\usepackage{epsfig,latexsym,cancel,amssymb,amsmath,verbatim,mathrsfs}
\usepackage{color}
\usepackage{graphicx}

\def\ra{\rightarrow}
\def\L{\left(}
\def\R{\right)}
\def\wt{\widetilde}

\def\ld{\lambda}
\def\f{\frac}
\newcommand{\be}{\begin{equation}}
\newcommand{\ee}{\end{equation}}
\newcommand{\bea}{\begin{eqnarray}}
\newcommand{\eea}{\end{eqnarray}}
\newcommand{\ba}{\begin{array}}
\newcommand{\ea}{\end{array}}

\long\def\symbolfootnote[#1]#2{\begingroup%
\def\thefootnote{\fnsymbol{footnote}}\footnote[#1]{#2}\endgroup}

\newcommand{\beq}{\begin{equation}}
\newcommand{\eeq}{\end{equation}}
\begin{document}

\title{Probing the CP-even Higgs Sector via $H_3\to H_2H_1$ in the Natural NMSSM}

\author{Zhaofeng Kang}
\email[E-mail: ]{zhaofengkang@gmail.com}
\affiliation{Center for High-Energy Physics, Peking University, Beijing, 100871, P. R. China}

\author{Jinmian Li}
 \email[E-mail: ]{jmli@itp.ac.cn}
\affiliation{State Key Laboratory of Theoretical Physics,
      Institute of Theoretical Physics, Chinese Academy of Sciences,
Beijing 100190, P. R. China}

\author{Tianjun Li}
\email[E-mail: ]{tli@itp.ac.cn}
\affiliation{State Key Laboratory of Theoretical Physics,
      Institute of Theoretical Physics, Chinese Academy of Sciences,
Beijing 100190, P. R. China}

\affiliation{School of Physical Electronics,
University of Electronic Science and Technology of China,
Chengdu 610054, P. R. China}

\affiliation{George P. and Cynthia W. Mitchell Institute for 
Fundamental Physics and Astronomy,
Texas A\&M University, College Station, TX 77843, USA }

\author{Da Liu}
\email[E-mail: ]{liudaphysics@gmail.com}
\affiliation{State Key Laboratory of Theoretical Physics,
      Institute of Theoretical Physics, Chinese Academy of Sciences,
Beijing 100190, P. R. China}

\author{Jing Shu}
\email[E-mail: ]{shujingtom@gmail.com}
\affiliation{State Key Laboratory of Theoretical Physics,
      Institute of Theoretical Physics, Chinese Academy of Sciences,
Beijing 100190, P. R. China}

\date{\today}

\begin{abstract}

After the discovery of a Standard Model (SM) like Higgs boson, naturalness strongly
favors the next to the Minimal Supersymmetric SM (NMSSM).  In this letter, we point out that 
the most natural NMSSM predicts the following CP-even Higgs $H_i$ sector: 
(A) $H_2$ is the SM-like Higgs boson with mass pushed-upward
by a lighter $H_1$ with mass overwhelmingly within $[m_{H_2}/2,m_{H_2}]$; 
(B) $m_{H_3}\simeq 2\mu/\sin2\beta\gtrsim300$ GeV;
(C) $H_3$ has a significant coupling to top quarks and can decay to $H_1H_2$ with
a large branching ratio. Using jet substructure we show that all the three 
Higgs bosons can be discovered via $gg\to H_3 \to H_1H_2\to b\bar b \ell\nu jj$ 
at the 14 TeV LHC. Especially, the
LEP-LHC scenario with $H_1\simeq98$ GeV has a very good discovery potential.

\end{abstract}

\maketitle

\noindent {\bf{Introduction:}} Supersymmetry provides the most elegant solution
to the gauge hierarchy problem in the Standard Model (SM). In the supersymmetric
SMs (SSMs) with $R$-parity, we can not only achieve the gauge coupling
unification, but also have a cold dark matter candidate. Recently,
the discovery of a SM like Higgs boson at the LHC with mass $m_h$ around 
126 GeV~\cite{LHC:Higgs} has deep implications to the SSMs.  Although such a
relatively heavy Higgs boson mass can be achieved in the Minimal
SSM (MSSM), it generically incurs a large fine-tuning 
(For the possible solutions, see~\cite{Antusch:2012gv}). 
By constrast, the next-to-the MSSM (NMSSM) with an extra SM singlet Higgs field $S$
is strongly favored by naturalness~\cite{Ellwanger:2009dp}, due to originally its dynamically
solution to the Higgs bilinear mass $\mu$ problem and now the SM-like Higgs boson mass enhancement 
via the relatively large Higgs trilinear Yukawa coupling $\ld$ in the superpotential 
and singlet-doublet mixing effect~\cite{Kang:2012sy,Cao:2012fz,Agashe:2012zq,Chang:2005ht}. 
The natural NMSSM may leave hints at the light stop sector, but the search is rather model 
dependent~\cite{Cao:2012rz,Bi:2012jv} and barely has relation with Higgs 
sector (Recent attempt to search for the light stop utilizing the properties of 
the SM-like Higgs boson was done in~\cite{Berenstein:2012fc}.). 

In the natural NMSSM, the second lightest CP-even Higgs boson $H_2$  is indentified
as the SM like Higgs boson, while the lightest CP-even Higgs boson 
 $H_1$ has dominant singlet component. Thus, the $H_2$ mass can be
\textbf{pushed-upward} via the singlet-doublet mixing 
effect~\cite{Kang:2012sy,Cao:2012fz,Agashe:2012zq,Chang:2005ht}. 
Such a scenario can explain the possible di-photon excess
from Higgs decays~\cite{Kang:2012sy,Ellwanger,Choi:2012he}
since the significant mixing effect reduces the decay width of $H_2\ra b\bar b$
and the light charged Higgsino may increase the Higgs decays to diphotons.
Interestingly, $H_1$ may be used to interpret the slight LEP
excess for the Higgs mass around 98 GeV~\cite{Barate:2003sz} (It receives some 
interest~\cite{Belanger:2012tt,Drees:2012fb} recently.), or the LHC excess for the Higgs mass 
around $\sim$113 GeV~\cite{cms113h}. 
A scenario with two light higgs and a low-mass pseudoscalar in NMSSM has been discussed in \cite{Cerdeno:2013cz}. 
More noticeable features emerge when we take 
the heavy CP-even Higgs boson $H_3$ into account. 
In this letter, we consider the CP-even Higgs sector in 
the natural NMSSM. We point out that
naturalness implies the $H_3$ mass range $m_{H_3}\in [300,600]$ GeV and 
its significant triple Higgs coupling with $H_1$ and $H_2$.
Such a Higgs sector structure leads us to investigate the discovery potential of
the whole CP-even Higgs bosons from the process $gg\ra H_3\ra H_1H_2$. 
With jet substructure,  we show that all three CP-even Higgs bosons $H_i$ can be probed 
at the 14 TeV LHC. Our search strategy is specially suitable for
the LEP-LHC Higgs bosons but also applies to the general pushing-upward scenario.

\noindent {\bf Light Higgs Bosons in the Pushing-Upward Scenario:} The SM-like Higgs boson 
can be accommodated without recurring severe fine-tuning, and 
we can show that the whole Higgs sector is light. Restricted to the $Z_3-$NMSSM, 
naturalness conditions point to a predictive parameter space
\begin{align}
&\ld:~0.6-0.7, \quad \tan\beta:~1.3-3.0,\cr
&\mu=\ld v_s:~100 {\rm \,GeV}-200{\rm \,GeV},
\end{align}
where $\tan\beta$ is the ratio of the vacuum expectation values for two Higgs
doublets, and $\kappa$ is the singlet cubic coupling in the superpotential. 
Also, $\kappa$ is constrained by perturbativity, and typically is no more than
half of $\ld$. The stop sector should be sufficiently light, e.g., 
$m_{\wt t_L}=m_{\wt t_R}=500$ GeV, and a flavor safe choice $A_t=-500$ GeV. 
Their concrete values will not qualitatively affect our following discussions.

Importantly, $A_\ld$ can be further determined in the
pushing-upward mixing scenario.  The Higgs mass square matrix
in the Goldstone basis is
{\small\begin{align}\label{Higgs:even}
&(M_S^2)_{11}=M_A^2+(m_Z^2-\ld^2v^2)\sin^22\beta,\cr
& (M_S^2)_{12}=-\f{1}{2}(m_Z^2-\ld^2v^2)\sin4\beta,\cr
&(M_S^2)_{13}=-\frac{1}{2}(M_A^2\sin2\beta+2\ld\kappa v_s^2)\cos2\beta\f{v}{v_s},\cr
&(M_S^2)_{22}=m_Z^2\cos^22\beta+\ld^2v^2\sin^22\beta,\cr
&(M_S^2)_{23}=\f{1}{2}(4\ld^2v_s^2-M_A^2\sin^22\beta-2\ld\kappa v_s^2\sin2\beta)\f{v}{v_s},\cr
&(M_S^2)_{33}=\f{1}{4}M_A^2\sin^22\beta\L\f{v}{v_s}\R^2\cr
&+4\kappa^2v_s^2+\kappa A_\kappa v_s-\f{1}{2}\ld\kappa v^2\sin2\beta~,~\,
\end{align}}
where $M_A^2=2\ld v_s(A_\ld+\kappa v_s)/\sin2\beta$ defines the largest scale
among these elements. Let the orthogonal matrix diagonalizing $M_S^2$ be $O$: 
$O^T{\rm Diag}(m_{H_3}^2,m_{H_2}^2,m_{H_1}^2)O=M_S^2$. 
The singlet-doublet mixing effect can be approximately studied by decoupling
the entries involving the first state. Ref.~\cite{Kang:2012sy} found that,
in the case with a large $\ld$ and small $\mu$, the realization of pushing-upward
scenario, which requires $(M_S^2)_{33}\lesssim(M_S^2)_{22}$, necessitates
a cancelation to reduce the large non-diagonal element $(M_{S}^2)_{23}$:
\begin{align}
1-\L A_\ld/2\mu+\kappa/\ld\R\sin2\beta\simeq 0.
\end{align}
Thus, $A_\ld$ is largely determined by $\mu$ and $\tan\beta$, and
to a less degree, by $\kappa$. Then we have
\begin{align}
m_{H_3}^2\approx M_A^2\simeq \L\f{2\mu}{\sin2\beta}\R^2\L1-\f{\kappa}{\ld}\f{\sin2\beta}{2}\R.
\end{align}
Recall that $\kappa<\ld$, so, to a good approximation, we get $m_{H_3}\simeq M_A\simeq 2\mu/\sin2\beta$,
which is about $2.5\mu$, relating the $H_3$ mass directly with the weak scale naturalness.

We now summarize the Higgs spectra in the natural NMSSM under consideration. First,
all the Higgs fields are properly light. $H_3$ and its $SU(2)_L$
partners, the charged Higgs bosons $H^\pm$ and the heavy CP-odd Higgs $A_2$, take roughly
degenerate masses $M_A$. $H_2$ is SM-like while $H_1$ is even lighter.
 $H_1$ is a SM singlet like and then can be allowed by the LEP experiment. Note that 
$m_{H_1}$ is most likely to fall into the region $[m_h/2,m_h]$ with the lower bound set
 by forbidding the decay $H_2\ra H_1H_1$ (Ref.~\cite{King:2012tr} considered such case.).
Otherwise it tends to be the dominant decay mode of $H_2$. 
In addition, the lightest CP-odd Higgs boson $A_1$ also has a mass 
around the weak scale. Moreover, a pair of charginos and three neutralinos, consisting 
of the Higgsinos and singlino, are light as well. 
All of them may be detectable at the LHC and here we focus on the CP-even Higgs bosons.

\noindent {\bf{$H_i-$couplings:}}
The Higgs signals at colliders are sensitive to their mixing angles whose effects, 
in a standard form, are described by the tree-level Lagrangian:
 \begin{align}\label{tree}
{\cal L}_{\rm tree}\supset &
r_{i,Z}\f{M_Z^2}{\sqrt{2}v}H_iZZ+r_{i,W}\f{\sqrt{2}M_W^2}{v}H_iW^+W^-\cr
&-r_{i,f}\f{m_f}{\sqrt{2}v}H_i\bar ff
%-r_{i,\phi}\f{m_\phi^2}{\sqrt{2}v}H_i\phi^\dagger \phi
+\mu_{ijk}H_iH_jH_k,
\end{align}
with $v\approx174$ GeV. %Here $f$ denotes any Dirac fermion. 
$r_{i,V}$. etc., encode the
deviations of $H_i$ from $h_{\rm SM}$. For instance, we have
\begin{align}\label{HVV}
r_{1,V}=O_{32},\quad r_{2,V}=O_{22},\quad r_{3,V}=O_{12}.
\end{align}
We also include the triple Higgs couplings, which will play a crucial
role in the search for Higgs bosons.

We now present the features of $H_3$ couplings. Firstly, note that
$(M_{S}^2)_{12}$ is a small entry and we can express it in terms of
$O$ and $m_{H_i}^2$. Since $m_{H_3}$ is a few times of $m_{H_{2,1}}$,
then it is not difficult to obtain the upper bound
\begin{align}
O_{12}=-s_{\theta_1}\lesssim (M_S^2)_{12}/m_{H_3}^2\sim (M_S^2)_{12}/(M_S^2)_{11},
\end{align}
where $(M_S^2)_{11}$ gives the dominant contribution to $m_{H_3}$.
Therefore, the trilinear couplings between $H_3$ and the weak gauge bosons are
negligibly small. Next, the reduced couplings of $H_3$ to the bottom and
top quarks are given by
\begin{align}
C_{3,b}=&-O_{11}\tan\beta +O_{12}\approx -O_{11}\tan\beta,\cr
 C_{3,t}=& O_{11}\cot\beta+O_{12}\approx O_{11}\cot\beta.
\end{align}
Owing to a relatively small $\tan\beta$ in the natural NMSSM, $H_3$
coupling to the bottom quark is not enhanced while its coupling
to the top quark is significant. They have crucial implications to
the collider phenomenology of $H_3$, e.g., it can be considerably
produced at the LHC by virtue of the significant coupling
to gluons:
\begin{align}
C_{3,g}=&1.03C_{2,t}-0.06C_{2,b}\approx O_{11}\cot\beta.
\end{align}
Finally, the triple Higgs coupling $H_3H_2H_1$ receives two possible large
contributions and is given by
\begin{align}
\mu_{123}\sim-\f{\ld A_\ld}{\sqrt{2}}\L 1+2\f{\kappa}{\ld}\f{\mu}{A_\ld}\R\simeq-\f{\ld A_\ld}{\sqrt{2}}.
\end{align}
It thus has a large $\ld A_\ld$ enhancement and leads to
$H_3\ra H_1H_2$ decay width 
%\begin{align}\label{H_321}
%\Gamma(H_3\ra H_1H_2)\simeq \f{\L\mu_{123}/m_{H_3}\R^2}{8\pi}m_{H_3}.
%\end{align}
at the GeV scale and dominates the $H_3$ Higgs-to-Higgs decay, as 
provides the most promising discovery prospect for 
$H_3$ and $H_1$, similarly to Ref.~\cite{Dolan:2012ac}.

We now turn our attention to the lightest Higgs boson $H_1$. Interestingly, 
the LEP collaboration reported (with an signal significance 2.3 $\sigma$)
a slight excess of events for a Higgs boson with mass $\sim 95-100$
GeV~\cite{Barate:2003sz}. Although our discussions on the Higgs bosons 
and the ensuing search strategy are not restricted to this case, 
it is tempting to interpret $H_1$ as the source of
this excess. So we have
\begin{align}
C_{1,V}^2\frac{{\rm Br}(H_1\ra b\bar b)}{{\rm Br}_{SM}(H_1\ra b\bar b)}\sim 0.1-0.25.
\end{align}
For $m_{ H}\lesssim100$ GeV, its decay to $b\bar b$ nearly
determines its total width. Thus, the LEP requires $C_{H_1,VV}\sim
0.3$ which is a typical value expected from the mixing Higgs sector.
\begin{figure}[htb]
\begin{center}
\includegraphics[width=2.5in]{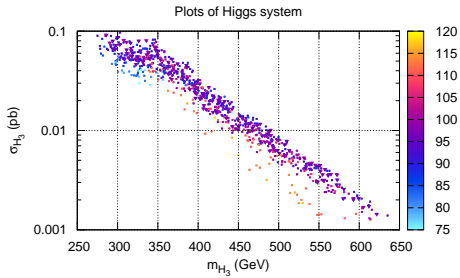}
\end{center}
\caption{\label{LEP-LHC} A plot on the $\sigma_{H_3}$-$m_{H_3}$ plane, with 
color code denoting $m_{H_1}$. Large inverted triangle points satisfy the LEP-LHC scenario. 
We use NMSSMTools 3.2.1~\cite{NMSSMTools}, set 
$ \f{A_\ld}{\rm GeV}\,\subset[300,\,500]$, $\f{A_\kappa}{\rm GeV}\subset[-300,\,0]$,
and require $\f{m_{H_2}}{\rm GeV}\subset[125,127]$ and signal strengthes 
$R_{2,gg}(\gamma\gamma)\subset[1.4,1.6]$, $R_{2,gg}(VV)\subset[1.0,1.3]$.}
\end{figure}

\noindent {\bf{Signature and backgrounds}:} In light of the previous analysis,
the signature $gg\ra H_3\ra H_1(\ra b\bar b) H_2(\ra W_hW_\ell)$ is promising, where we denote $W_h$ as hadronic decaying $W$ boson and denote $W_\ell$ as leptonic decaying $W$ boson.
The $W_\ell$ will suppress the enormous QCD backgrounds. The total cross
section is
{\small\begin{align}
\sigma_{H_3}=& 0.2\L\f{C_{3,g}}{0.4}\R^2 \f{{\rm Br}(H_3\ra H_1H_2)}{20\%}
\f{{\rm Br}(H_1\ra b\bar b)}{90\%}\cr
&
\f{{\rm Br}(H_2\ra W_\ell W_h)}{28\%}\f{ \sigma_{\rm GF}(h_{\rm SM})}{10{\,\rm pb}}\,\rm pb,
\end{align}}where $\ell=e,\mu$. The numerical results are shown in Fig.~\ref{LEP-LHC}, where a plot of
the distribution of $\sigma_{H_3}$ on the $m_{H_2}-m_{H_1}$ plane is presented.
It can be seen that its values cluster well for a given $m_{H_3}$
(typically within only a few times), in particular for heavier $H_3$.

We implement the simplified model for Higgs bosons in Feynrules~\cite{Feynrule} to
generate the UFO format of the effective model for MadGraph5~\cite{Alwall:2011uj},
where the parton-level signatures are generated.

The semi-leptonic $t\bar{t}$ pair production is the dominant background (BG),
with the NNLO cross section $\approx 240$ pb~\cite{Ahrens:2011px}. The subdominant BG
$W_\ell+b\bar b+$jets has cross section depending on the renormalization scale, roughly, about $40$ pb.
Other backgrounds can be neglected in our signal region. BGs are generated
using MadGraph5. To avoid double counting, we adopt the modified version of
MLM-matching~\cite{Hoche:2006ph} with $xq{\rm cut}=15$ GeV. For the latter BG, we include
up to 2 additional jets and set the $k-$factor to be 2.

We use PYTHIA6.420~\cite{Sjostrand:2006za} for decaying particles, parton-showering
and hadronization. However, in order to employ the BDRS
procedure later, we turn off the $B-$hadron decays in Pythia. The produced objects
are then converted to the HepMC~\cite{Dobbs:2001ck} event format and passed to
Fastjet 3.0~\cite{Cacciari:2011ma} to cluster the final states. The final visible
particles are requited to have $p_T >$ 0.1 GeV and $|\eta| < 5.0$ which are defined
as tracks hereafter. Leptons from signal events should be isolated, otherwise they
are combined with the tracks to reconstruct fat jets later. Additionally,
signal leptons are required to have $|\eta|< 2.5 $ and $p_T > 10$ GeV. We take $b-$tagging
efficiency of 70$\%$ with the other light quark mis-tagged probability 1$\%$.

We choose the C/A algorithm~\cite{C/A} with radius $R$=$1.4$
and $p_T$$ >$40 GeV to cluster the tracks and form fat jets.
Following BDRS~\cite{Butterworth:2008iy}, we first break the hard fat jets %with $p_T > $ 40 GeV 
into subjets $j_{1,2}$ with masses $m_{j_{1,2}}$. % are ordered as $m_{j_1} > m_{j_2}$ by convention.
Next, a significant mass drop $m_{j_1}$ $<$ $\mu\, m_{j}$ with $\mu$=0.667 and not 
too asymmetric splitting, i.e., $y=$min($p_{T,j_1}^2,\,\,p_{T,j_2}^2)\Delta R^2_{j_1,j_2}/m_j^2 > y_{cut}$
with $y_{cut}$=0.09 ($\Delta R^2_{j_1,j_2}$ is the angular distance), are required. If the above criterion are not satisfied, we will set $j=j_1$ and go back to decomposition.  
Finally, we filter the Higgs neighbourhood,  % as done in the original BDRS method:
resolving the fat jets on a finer angular scale $R_{\text{filt}}=
\text{min}(0.35, R_{j_1j_2}/2)$ and taking the three hardest objects, with the remains identified as
the underlying events contamination and hence dropped.

\noindent {\bf{Events selection and results}:}
Two basic cuts are imposed to trigger our events. Firstly, at least two filtered fat jets
are required. One of them has two leading subjets which pass $b-$tagging and satisfy $|\eta|<2.5$,
and then is identified as the $H_1-$jet. Among the remaining fat jets, the one with highest $p_T$
is regarded as the $W_h-$jet~\cite{Papaefstathiou:2012qe}. Secondly, 
the events must contain exactly one isolated lepton.

\begin{figure}[htb]
\begin{center}
\begin{tabular}{cc}
\includegraphics[width=0.2\textwidth]{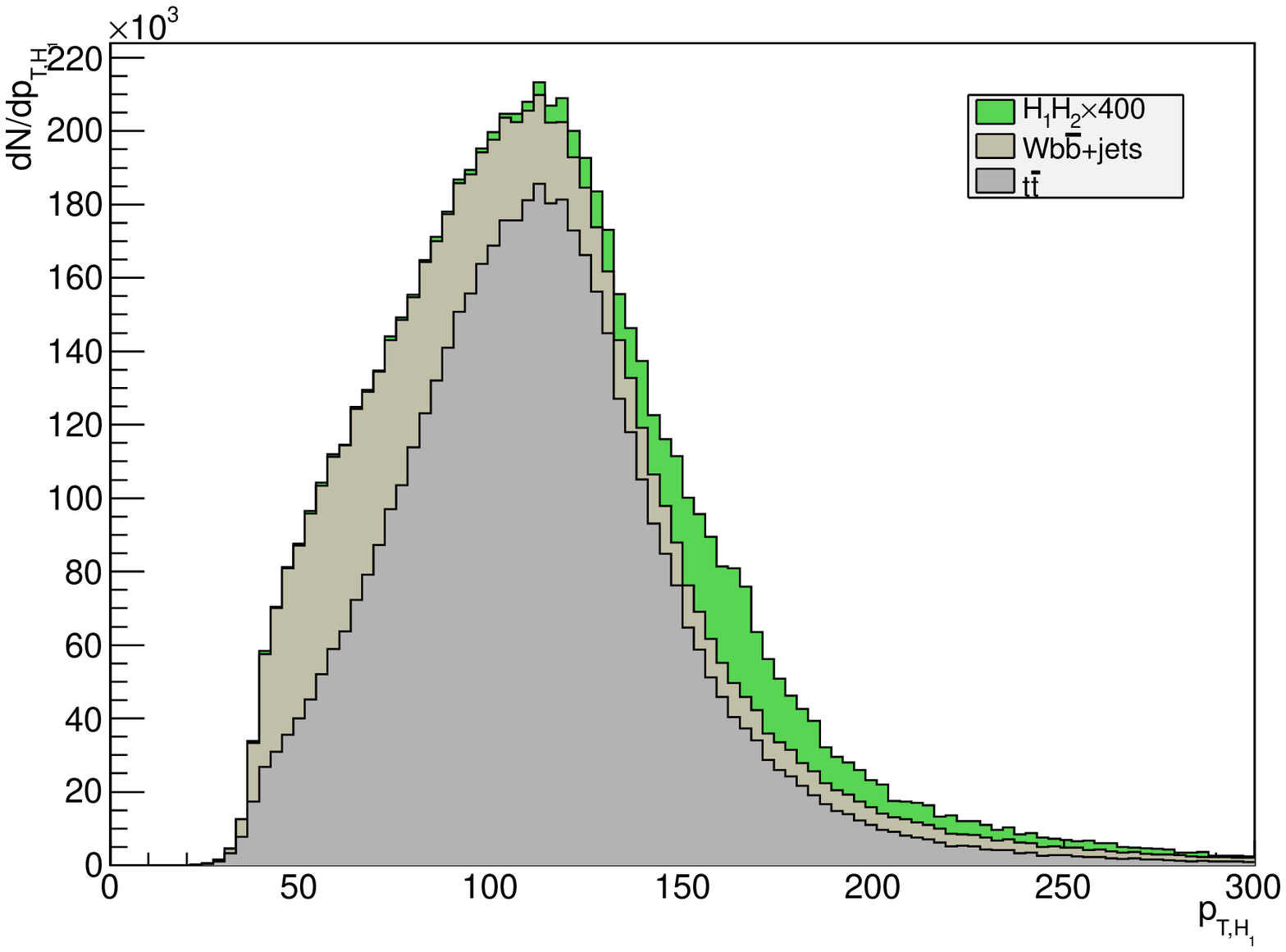} &
\includegraphics[width=0.2\textwidth]{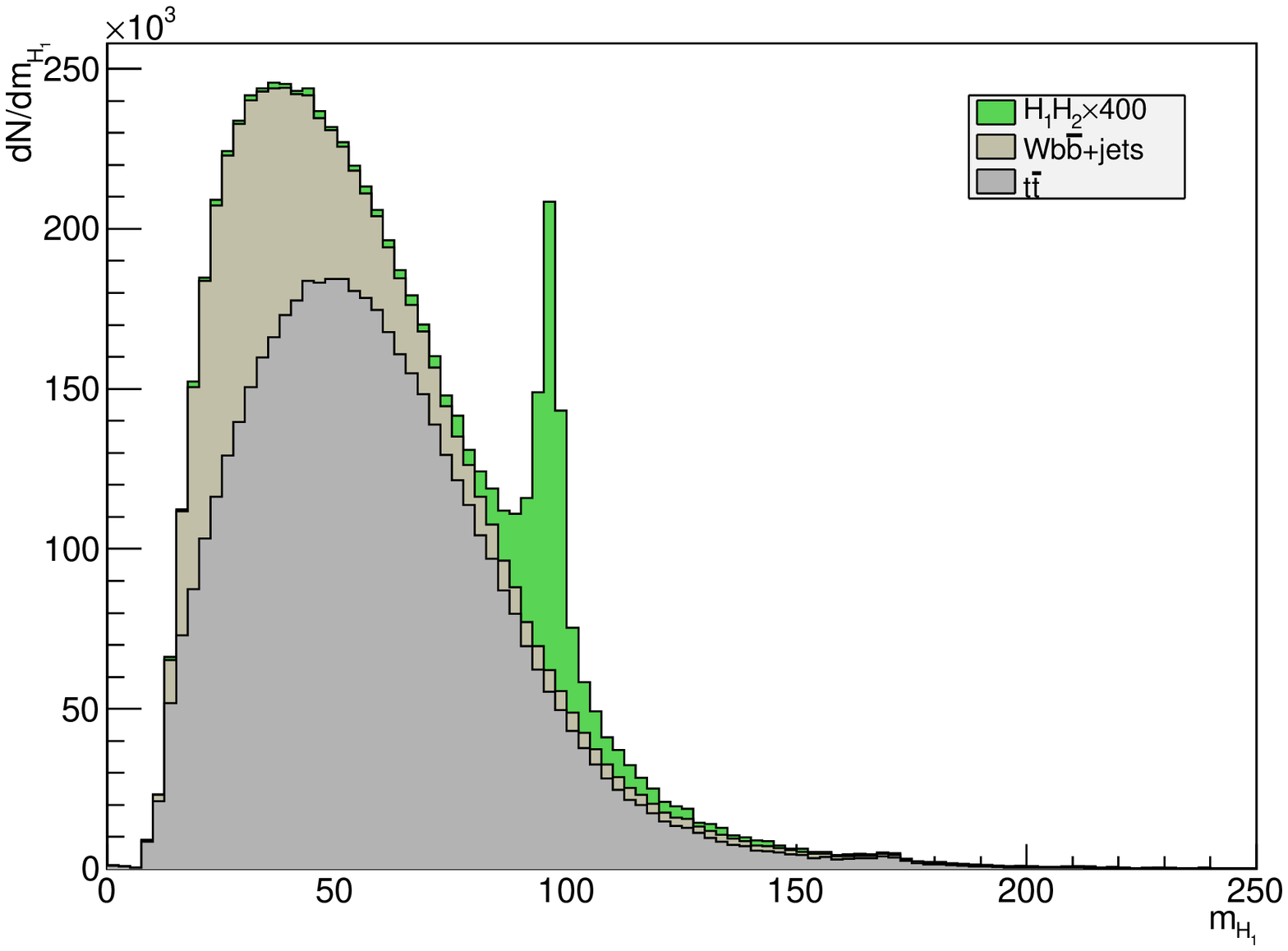}\\
\includegraphics[width=0.2\textwidth]{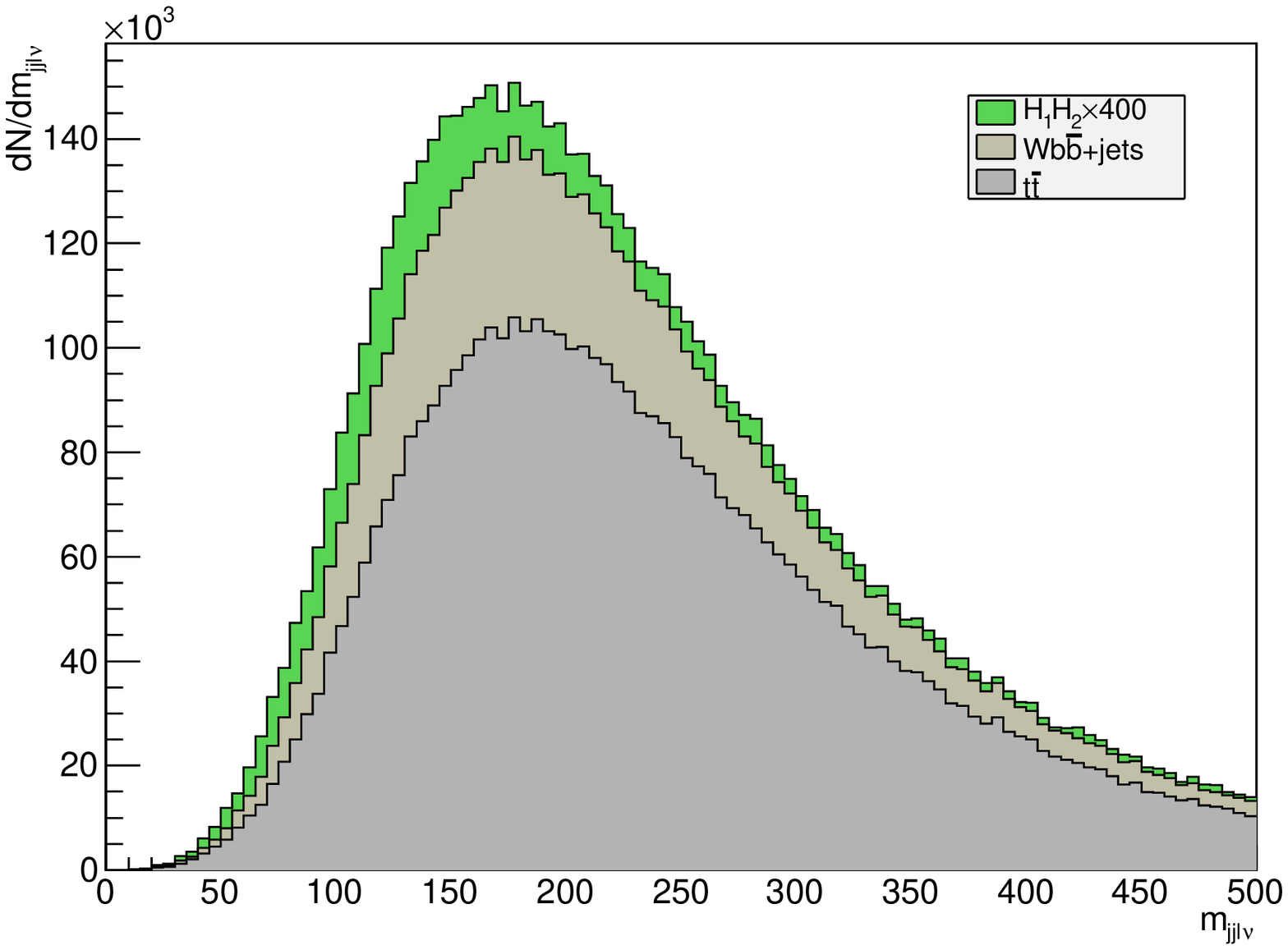}&
\includegraphics[width=0.2\textwidth]{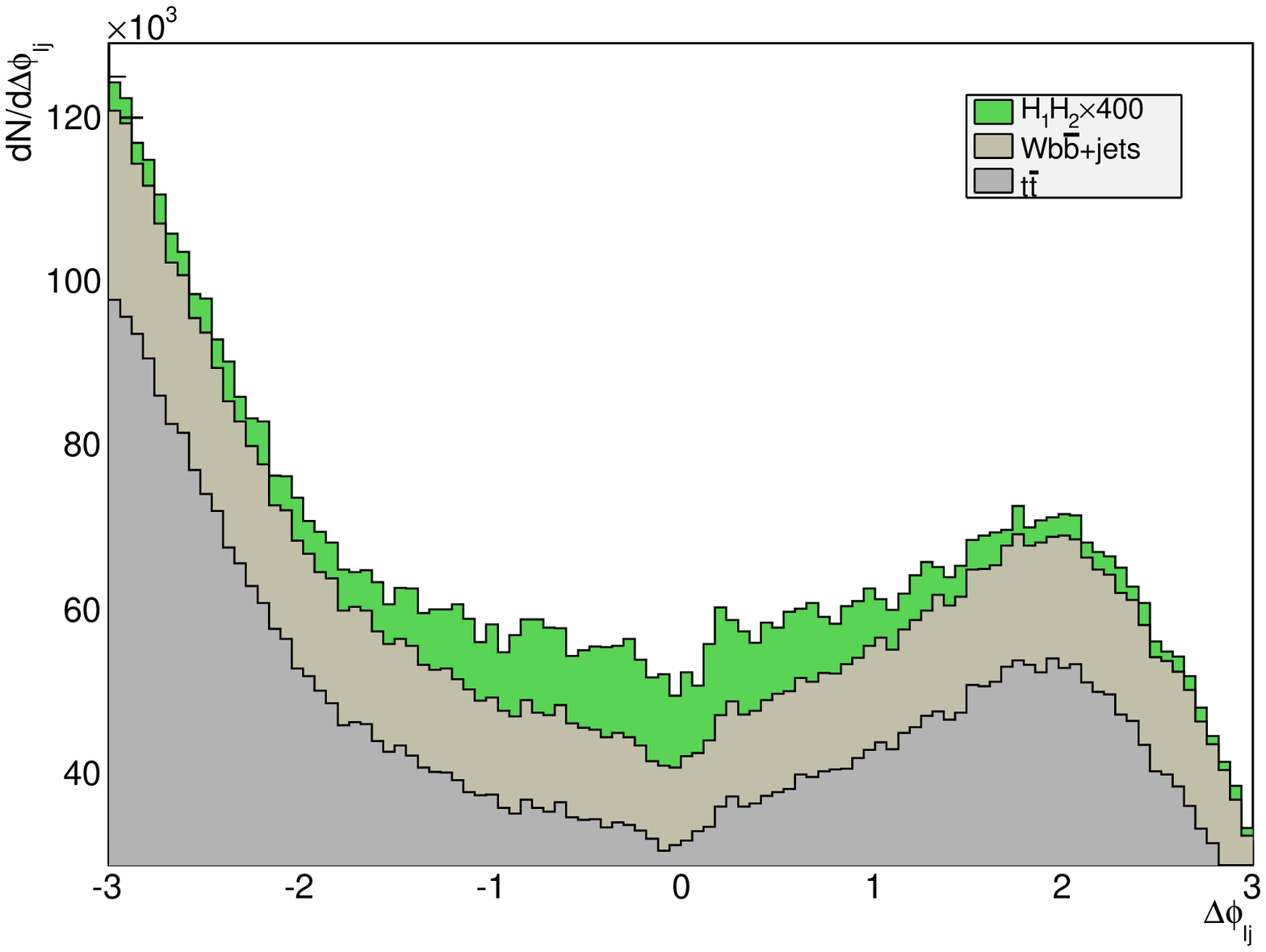}\\
\includegraphics[width=0.2\textwidth]{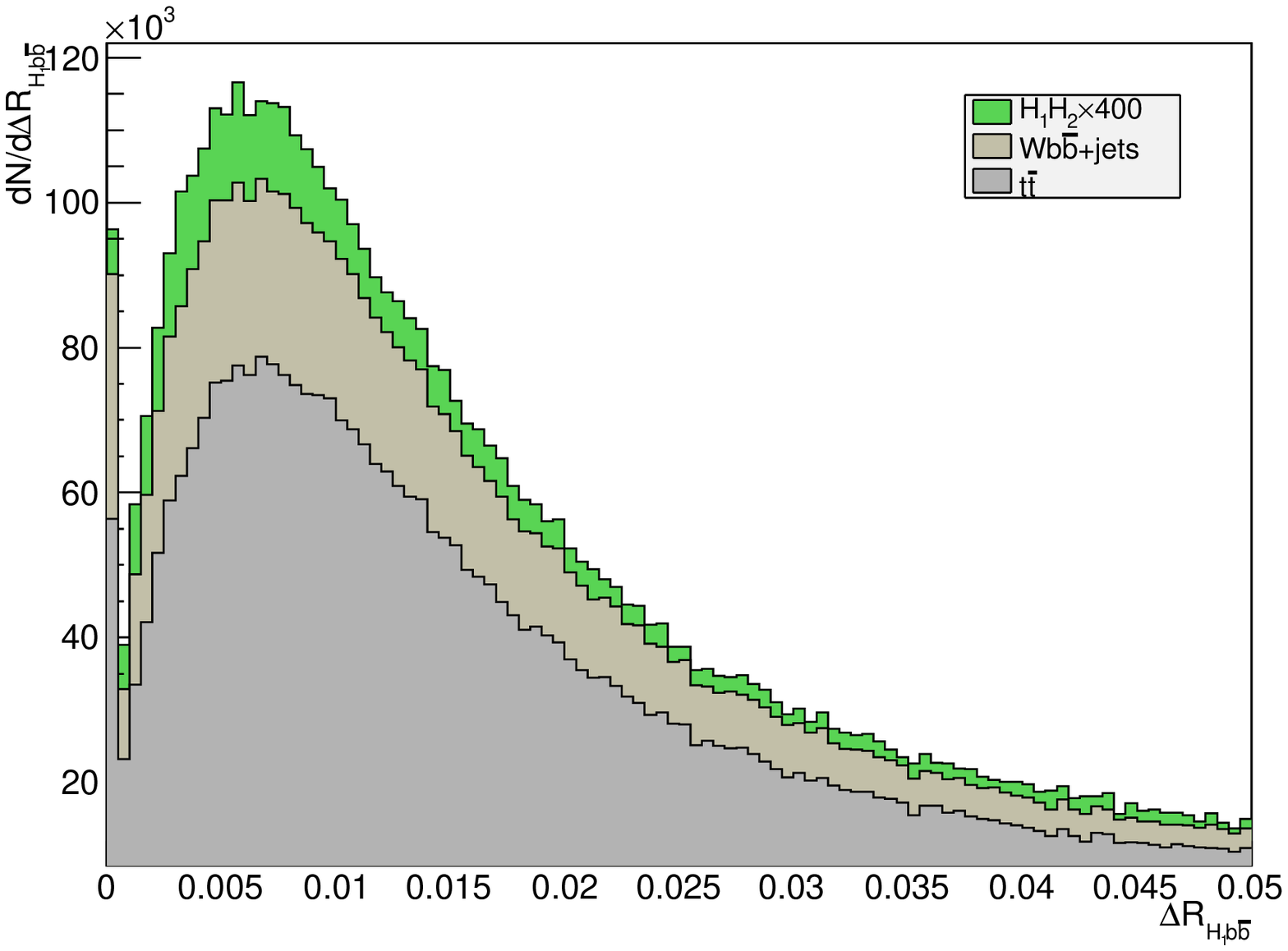}&
\includegraphics[width=0.2\textwidth]{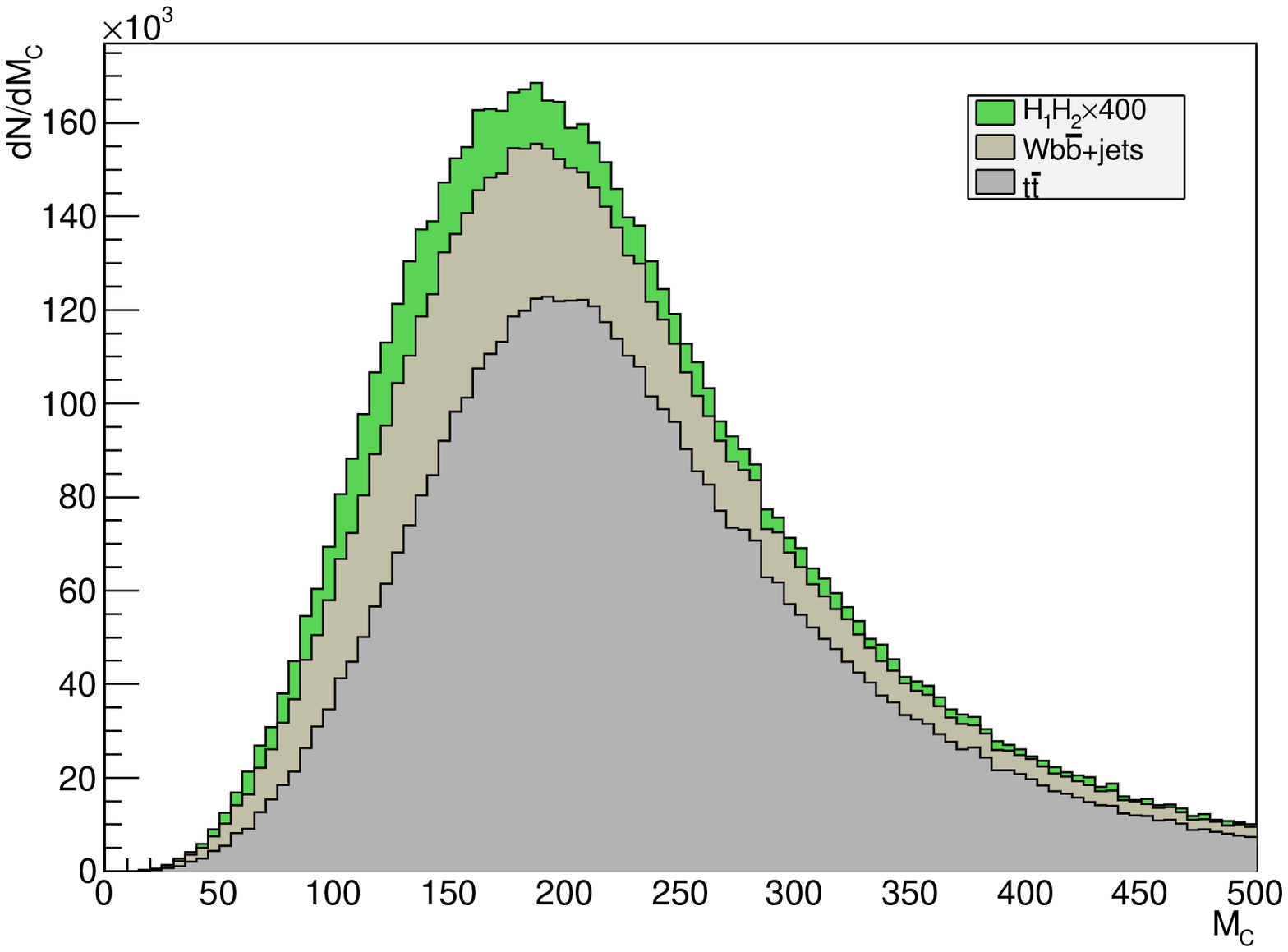}\\
\end{tabular}
\caption{Distribution for trigged signal and background of $p_{T,H_1}$, $m_{H_1}$, $m_{jjl\nu}$, $\Delta \phi_{lj}$, $\Delta R_{H_1b \bar{b}}$, $M_C$. The number of events have been normalised to 14 TeV  500 $fb^{-1}$, and the signal is 400 times amplified. }
\label{kins}
\end{center}
\end{figure}
%\subsection{Cut flow and kinematice distrubution}
%In the following,
For illumination, here we will take a benchmark point inspired by the LEP-LHC Higgs scenario:
$m_{H_1}=98 $GeV, $m_{H_2}=125$ GeV as well as $m_{H_3}=400$ GeV. Figure~\ref{kins} shows
the distributions of some important kinematic variables. In terms of the plots, we display
the cut flow:
\begin{itemize}
\item{Cut1:}
The relatively large mass splitting between $H_3$ and $H_1$ gives $H_1$ a boost.
Therefore, we require $p_{T,b\bar b}>150 {\rm\, GeV}$, $p_{T,jj\ell\, \nu}>120 {\rm\, GeV}$,
and $|p_{T,b\bar b}-p_{T,jj\ell\, \nu}|<20 \rm \,GeV$.
\item{Cut2:}  It is observed that the longitudinal momentum of the neutrino from $W$ decay
is generically small, and hence $m_{H_{2,3}}$ can be approximately reconstructed by assuming $p_{z,\nu}=0$. Practically, cuts based on this assumption are sufficiently good. So we impose:
$95 {\rm\, GeV}<m_{H_1}<100  {\rm\, GeV}$, $m_{jj \ell \nu}<150 {\rm\, GeV}$, and
$m_{b\bar b jj \ell \nu}<440 {\rm\, GeV}$.
\item{Cut3:} Because $H_2$ has spin-0 and $W$ only couples to the left-handed fermions,
the lepton from $W_\ell$ will align with one of the jets from $W_h$ decay. It allows
us to impose a cut $|\Delta \phi_{\ell j}|<1.5$, namely the azimuthal angles difference
between the signal lepton and (one) jet being sufficiently small.
\item{Cut4:} The filtered $H_1$-jet actually contains three subjets,
the $b\bar b$ and a radiated gluon. So the $H_1$-jet and its $b \bar{b}$
subsystem must have a very small angle distance. By contrast, the angle distance
between the $H_1$-jet and $W_h$-jet is much larger.
Thus, we require
$\Delta R_{H_1,b\bar{b}} < 0.01$, and $2.6<\Delta R_{H_1W_h}<3.4$.
\item{Cut5:} We also impose
 the cluster transverse mass of decay product of 
the $H_2$: $M_C=\sqrt{p^2_{T,jj\ell}+m^2_{jj\ell}}+\slashed E_T<220 $ GeV. 
\end{itemize}
With the above cuts, we obtain the signal significance 4.42 $\sigma$ excess
for the LEP-LHC benchmark point at 14 TeV 500 $fb^{-1}$.
The cut efficiency and the signals are presented in Table~\ref{cutflow}.
\begin{table}[htbp]
\centering
\begin{tabular}{c | c | c | c}
                            & $t\bar{t}$ & $W(\to l \nu j j )b\bar{b}+jets$ & Signal \\\hline
      Total              &  $1.2 \times10^8$   &   $1.91 \times 10^7$     &  $1.25 \times 10^4$ \\
       Trigged       &    $4.95 \times 10^6$ &  $1.45 \times 10^6$     &  1456.75\\
      Cut1           &      $3.77 \times 10^5$  & $1.61 \times 10^5$    &   639.5 \\
       Cut2            &     1932          &    203     &   119.75  \\
        Cut3            &    1512         &   155.2    &  105.5\\
        Cut4            &     108          &    47.75   & 56.25\\
        Cut5            &     84           &    47.75    & 55\\
\end{tabular}
\caption{Number of events after each cut for background and signal (normalized to 500 $fb^{-1}$).
The signal significance $S/\sqrt{S+B}$ has reached to 4.02 and with the precise
4.42 $\sigma$ excess for the LEP-LHC benchmark point. }
\label{cutflow}
\end{table}

Since Fig.~\ref{LEP-LHC} shows obvious cluster behavior, the whole parameter space with pushing-upward effect
can be explored. Using BDT analysis~\cite{bdt},
we consider six representative points to demonstrate the search
prospect, and the discovery signal significance for each case is given in Table~\ref{sixbdt}.
Some observations can be made: (A) For a given $m_{H_3}$, a lighter $H_1$ shows better discovery potential; 
(B) Increasing $H_3$ mass helps to boost $H_1$ but the cross section is reduced. Thus,  
a moderately heavy ${H_3}\sim 400$ GeV and relatively light $H_1$ have the most promising 
discovery potential; (C) Most of the parameter space is discoverable except for simultaneously 
light $H_3$ and heavy $H_1$, e.g., the benchmark point B1, despite of a rather large cross section, 
has  a quite low signal significance.
%For $m_{H_3}\simeq600$ GeV, event rate are small and thus
%the discovery potential is relatively low except for properly light $H_1$. Thus it is difficult to
%cover the whole space.

The situation can be further 
 improved when we take $H^\pm$ into account. $H^\pm$ can be produced
associated with a single top, with a moderately large cross section at the small $\tan\beta$ region.
Moreover, it can decay to $H_1$ and $W$ with a substantial branching ratio and hence
provide a way to probe $H_1$ and $H^\pm$. In this case $H^\pm$ can be produced with a
larger $p_T$ and the boost will be easier for lighter $H^\pm$, not very sensitive 
to $m_{H_1}$. So it can provide a complementary or even more promising 
channel for the pushing-upward scenario. We leave it for the future work.

\begin{table}[htbp]
\centering
\begin{tabular}{c | c | c | c | c}
                            & $m_{H_1}$(GeV) & $m_{H_3}$(GeV) & $\sigma$ (fb) & $\frac{S}{\sqrt{S+B}}$\\

      B1          &  100    &  300   & 70   &  0.81  \\
       B2            &  65  &  300   & 50 & 3.84 \\
             B3        &  98   & 400  & 25 &  4.73 \\
       B4      &   65 & 400 & 20 &  7.68 \\
        B5            &   100 &   600 & 2 & 2.79 \\
        B6            &    65    & 600  & 2 &  4.99\\
\end{tabular}
\caption{Discovery signal significances for 6 representative points at 14 TeV 500 $fb^{-1}$. We design 25 kinematic
variables for BDT analysis~\cite{bdt}: $\slashed E_T$,  $p_{T,W}$, $m_W$,
$n_{jet}$, $p_{T,b_1}$, $p_{T,b_2}$, $p_{T,\ell}$, $m_{T,\ell\nu}$,
$ p_{T,wj\ell}$, $p_{T,wj_2}$, $p_{T,jj\ell\nu}$, $\Delta R_{\ell j}$,
$\Delta \phi_{lw}$, $p_{T,H_3}$, $m_{\ell\nu}$, 
$E_{l\nu}$, and $m_{H_3}$. }
\label{sixbdt}
\end{table}

\noindent {\bf{Conclusion}:}
We pointed out the specific features in  the CP-even Higgs sector of 
the natural NMSSM, and  showed that all three CP-even Higgs boson $H_i$ can be probed 
at the 14 TeV LHC. 

\noindent {\bf{Acknowledgements}:}
We would like to thank Ran Huo, Chunli Tong, Andreas Papaefstathiou, Lilin Yang and Jose Zurita for helpful discussions.
This research was supported in part
by the Natural Science Foundation of China
under grant numbers 10821504, 11075194, 11135003, and 11275246,
and by the DOE grant DE-FG03-95-Er-40917 (TL).

\end{document}